\documentclass{iaus}
\usepackage{graphicx}

\pubyear{2012}
\volume{289}  
\jname{Advancing the Physics of Cosmic Distances}
\editors{Richard de Grijs \and Giuseppe Bono, eds.}
\newcommand{\apj}{{\it ApJ}}
\newcommand{\aj}{{\it AJ}}
\newcommand{\mnras}{{\it MNRAS}}
\newcommand{\aanda}{{\it A\&A}}
\newcommand{\pasp}{{\it PASP}}

\title[IAUS289.~~Statistical analysis of R$_{\it 0}$ determinations]
{Statistical analysis of the determinations\\ of the Sun's Galactocentric distance}
\author[Zinovy Malkin]{Zinovy Malkin$^{1,2}$}
\affiliation{$^1$Pulkovo Observatory, St. Petersburg 196140, Russia \\ $^2$St. Petersburg State University, St. Petersburg 198504, Russia \\
 email: {\tt malkin@gao.spb,ru}}

\begin{document}

\maketitle

\begin{abstract}
Based on several tens of R$_0$ measurements made during the past
decades, several studies have been performed to derive the best
estimate of R$_0$. Some used just simple averaging to derive a result,
whereas others provided comprehensive analyses of possible errors in
published results. In either case, detailed statistical analyses of
data used were not performed.  However, a computation of the best
estimates of the Galactic rotation constants is not only an
astronomical but also a metrological task.  Here we perform an
analysis of 53 R$_0$ measurements (published in the past 20 years) to
assess the consistency of the data.  Our analysis shows that they are
internally consistent.  It is also shown that any trend in the R$_0$
estimates from the last 20 years is statistically negligible, which
renders the presence of a bandwagon effect doubtful. On the other
hand, the formal errors in the published R$_0$ estimates improve
significantly with time.
\end{abstract}

\firstsection 
\section{Introduction}

Accurate knowledge of the distance from the Sun to the center of the
Galaxy, R$_0$, is important in many fields of astronomy and space
science.  In particular, the primary motivation for this study was a
wish to improve the accuracy of modeling the Galactic aberration
(Malkin 2011).

Over the past decades, many tens of R$_0$ determinations have been
made, making use of different principles and observing methods, and
thus characterized by different random and systematic
errors. Therefore, deriving a best R$_0$ estimate from these data is
not only an astronomical but also a metrological task, similar to
deriving the best estimates of the fundamental constants in physics.
For the latter, several statistical methods have been developed to
obtain best estimates---as well as their realistic
uncertainties---from heterogeneous measurements. In this study, we
applied those methods to the R$_0$ data.

Another goal of this study was to investigate a possible trend in the
multi-year series of R$_0$ estimates, as discussed by many authors.
However, estimates of any such trend differ significantly among
papers. Therefore, we tried to clarify this issue using the latest
results. More details are provided by Malkin (2012).

\section{Data used}

For the present study, we used all R$_0$ measurements published in the
period 1992--2011, with the exception of a number of results that were
revised in subsequent papers.  We did not use the results of Glushkova
et al. (1998; revised by Glushkova et al. 1999), Paczynski \& Stanek
(1998; revised by Stanek et al. 2000), Eisenhauer et al. (2003;
revised by Eisenhauer et al. 2005, which was in turn revised by
Gillessen et al. 2009). In total, 53 estimates (listed in
Table~\ref{tab:allr0}) were used.

Where both random (statistical) and systematic uncertainties were
given, they were summed in quadrature. If two different values were
given for the lower and upper boundaries of the confidence interval,
the mean value of these boundaries was used as the uncertainty in the
result (the lower and upper boundaries were close to each other in all
cases, so that this approximation procedure does not significantly
affect the final result). Where authors gave several estimates of
R$_0$ without a final preference, the unweighted average of these
estimates was computed.

\begin{table}
\centering
\caption{R$_0$ estimates used in this study.}
\label{tab:allr0}
{\scriptsize
\begin{tabular}{lll}
\hline \multicolumn{1}{c}{R$_0$ (kpc)} & \multicolumn{1}{c}{$\pm
  (1\sigma)$} & \multicolumn{1}{c}{Reference} \\ 
\hline 
7.9  & 0.8  & Merrifield, M.R. 1992, \aj, 103, 1552 \\ 
8.1  & 1.1  & Gwinn, C.R., et al. 1992, \apj, 393, 149 \\ 
7.6  & 0.6  & Moran, J.M., et al. 1993, \textit{Lect. Notes Phys.}, 412, 244 \\ 
7.6  & 0.4  & Maciel, W.J. 1993, \textit{Ap\&SS}, 206, 285 \\ 
8.09 & 0.3  & Pont, F., et al. 1994, \aanda, 285, 415 \\ 
7.5  & 1.0  & Nikiforov, I.I., \& Petrovskaya, I.V. 1994, \textit{Astron. Rep.}, 38, 642 \\ 
7.0  & 0.5  & Rastorguev, A.S., et al. 1994, \textit{Astron. Lett.}, 20, 591 \\ 
8.8  & 0.5  & Glass, I.S., et al. 1995, \mnras, 273, 383 \\ 
7.1  & 0.5  & Dambis, A.K., et al. 1995, \textit{Astron. Lett.}, 21, 291 \\ 
8.3  & 1.0  & Carney, B.W., et al. 1995, \aj, 110, 1674 \\ 
8.21 & 0.98 & Huterer, D., et al. 1995, \aj, 110, 2705 \\ 
7.95 & 0.4  & Layden, A.C., et al. 1996, \aj, 112, 2110 \\ 
7.55 & 0.7  & Honma, M., \& Sofue, Y. 1996, \textit{PASJ}, 48, L103 \\ 
8.1  & 0.4  & Feast, M.W. 1997, \mnras, 284, 761 \\ 
8.5  & 0.5  & Feast, M., \& Whitelock, P. 1997, \mnras, 291, 683 \\ 
7.66 & 0.54 & Metzger, M.R., et al. 1998, \aj, 115, 635 \\ 
8.1  & 0.15 & Udalski, A. 1998, \textit{Acta Astron.}, 48, 113 \\ 
7.1  & 0.4  & Olling, R.P., \& Merrifield, M.R. 1998, \mnras, 297, 943 \\ 
8.51 & 0.29 & Feast, M., et al. 1998, \mnras, 298, L43 \\ 
8.2  & 0.21 & Stanek, K.Z., \& Garnavich, P.M. 1998, \apj, 503, L131 \\ 
8.6  & 1.0  & Surdin, V.G. 1999, \textit{Astron. Astrophys. Trans.}, 18, 367 \\ 
7.4  & 0.3  & Glushkova, E.V., et al. 1999, \textit{Astron. Astrophys. Trans.}, 18, 349 \\ 
7.9  & 0.3  & McNamara, D.H. et al. 2000, \pasp, 112, 202 \\ 
8.67 & 0.4  & Stanek, K.Z., et al. 2000, \textit{Acta Astron.}, 50, 191 \\ 
8.2  & 0.7  & Nikiforov, I.I. 2000, \textit{Astron. Soc. Pac. Conf. Ser.}, 209, 403 \\ 
8.24 & 0.42 & Alves, D.R. 2000, \apj, 539, 732 \\ 
8.05 & 0.6  & Genzel, R., et al. 2000, \mnras, 317, 348 \\ 
8.3  & 0.3  & Gerasimenko, T.P. 2004, \textit{Astron. Rep.}, 48, 103 \\ 
7.7  & 0.15 & Babusiaux, C., \& Gilmore, G. 2005, \mnras, 358, 1309 \\ 
8.01 & 0.44 & Avedisova, V.S. 2005, \textit{Astron. Rep.}, 49, 435 \\ 
8.7  & 0.6  & Groenewegen, M.A.T., \& Blommaert, J.A.D.L. 2005, \aanda, 443, 143 \\ 
7.2  & 0.3  & Bica, E., et al. 2006, \aanda, 450, 105 \\ 
7.52 & 0.36 & Nishiyama, S., et al. 2006, \apj, 647, 1093 \\ 
8.1  & 0.7  & Shen, M., Zhu, Z. 2007, \textit{Chin. J. Astron. Astrophys.}, 7, 120 \\ 
7.4  & 0.3  & Bobylev, V.V., et al. 2007, \textit{Astron. Lett.}, 33, 720 \\ 
7.94 & 0.45 & Groenewegen, M.A.T., et al. 2008, \aanda, 481, 441 \\ 
8.07 & 0.35 & Trippe, S., et al. 2008, \aanda, 492, 419 \\ 
8.16 & 0.5  & Ghez, A.M., et al. 2008, \apj, 689, 1044 \\ 
8.33 & 0.35 & Gillessen, S., et al. 2009, \apj, 692, 1075 \\ 
8.7  & 0.5  & Vanhollebeke, E., 2009, \aanda, 498, 95 \\  
7.58 & 0.40 & Dambis, A.K. 2009, \mnras, 396, 553 \\ 
7.2  & 0.3  & Bonatto, C., et al. 2009, \textit{Globular Clusters: Guides to Galaxies}, p. 209 \\ 
8.4  & 0.6  & Reid, M.J., et al. 2009, \apj, 700, 137 \\ 
7.75 & 0.5  & Majaess, D.J., et al. 2009, \mnras, 398, 263 \\ 
7.9  & 0.75 & Reid, M.J., et al. 2009, \apj, 705, 1548 \\ 
8.24 & 0.43 & Matsunaga, N., et al. 2009, \mnras, 399, 1709 \\ 
8.28 & 0.33 & Gillessen, S., 2009, \apj, 707, L114 \\ 
7.7  & 0.4  & Dambis, A.K. 2010, \textit{Variable Stars, the Galactic halo and Galaxy Formation}, p. 177 \\ 
8.1  & 0.6  & Majaess, D. 2010, \textit{Acta Astron.}, 60, 55 \\ 
8.3  & 1.1  & Sato, M., et al. 2010, \apj, 720, 1055 \\ 
7.80 & 0.26 & Ando, K., et al. 2011, \textit{PASJ}, 63, 45 \\ 
8.3  & 0.23 & Brunthaler, A., et al. 2011, \textit{Astron. Nachr.}, 332, 461 \\ 
8.29 & 0.16 & McMillan, P.J. 2011, \mnras, 414, 2446 \\ 
\hline
\end{tabular}
}
\end{table}

\section{Results}

\begin{figure}[t]
\begin{minipage}[t]{0.5\linewidth}
\centering
\includegraphics[clip,width=\textwidth]{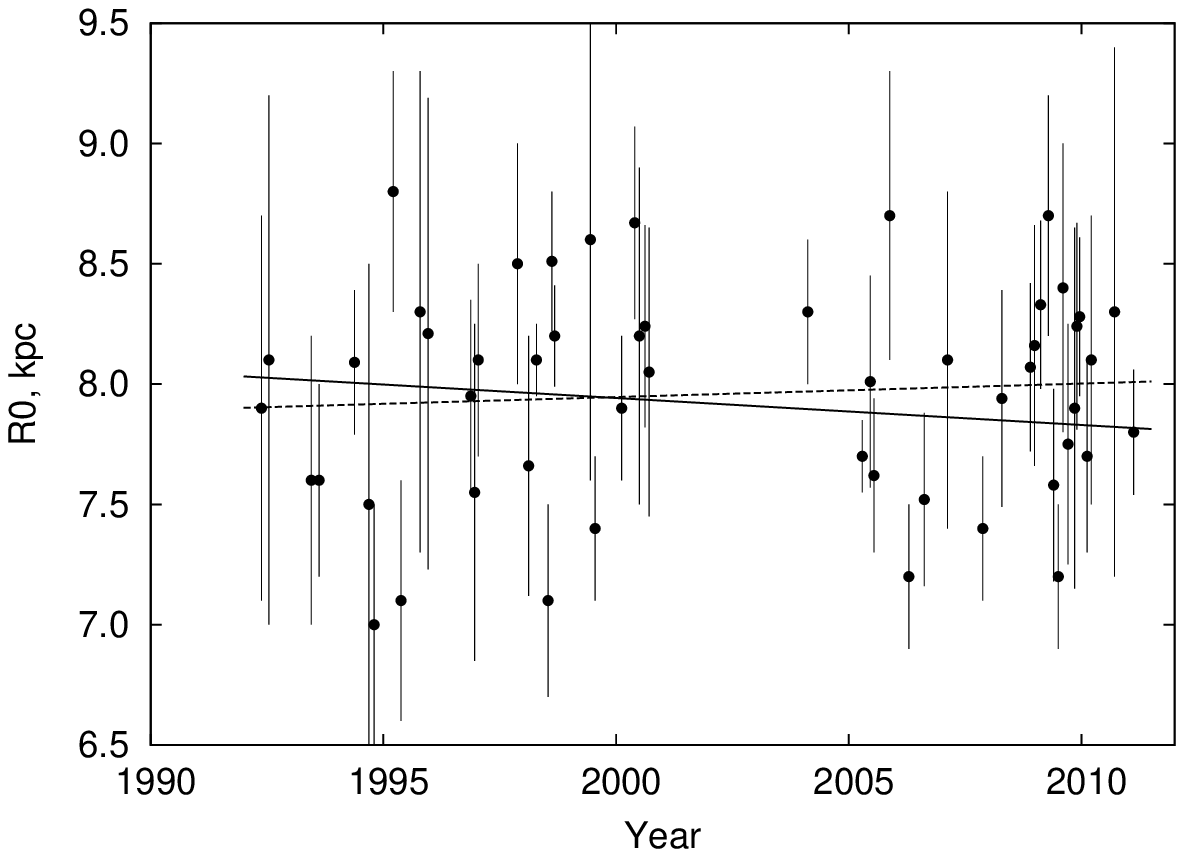}
\caption{R$_0$ estimates used in this study. The weighted (solid line)
  and unweighted (dashed line) trends are also shown.}
\label{fig:data_R0}
\end{minipage}
\hspace{0.01\textwidth}
\begin{minipage}[t]{0.5\textwidth}
\centering
\includegraphics[clip,width=\textwidth]{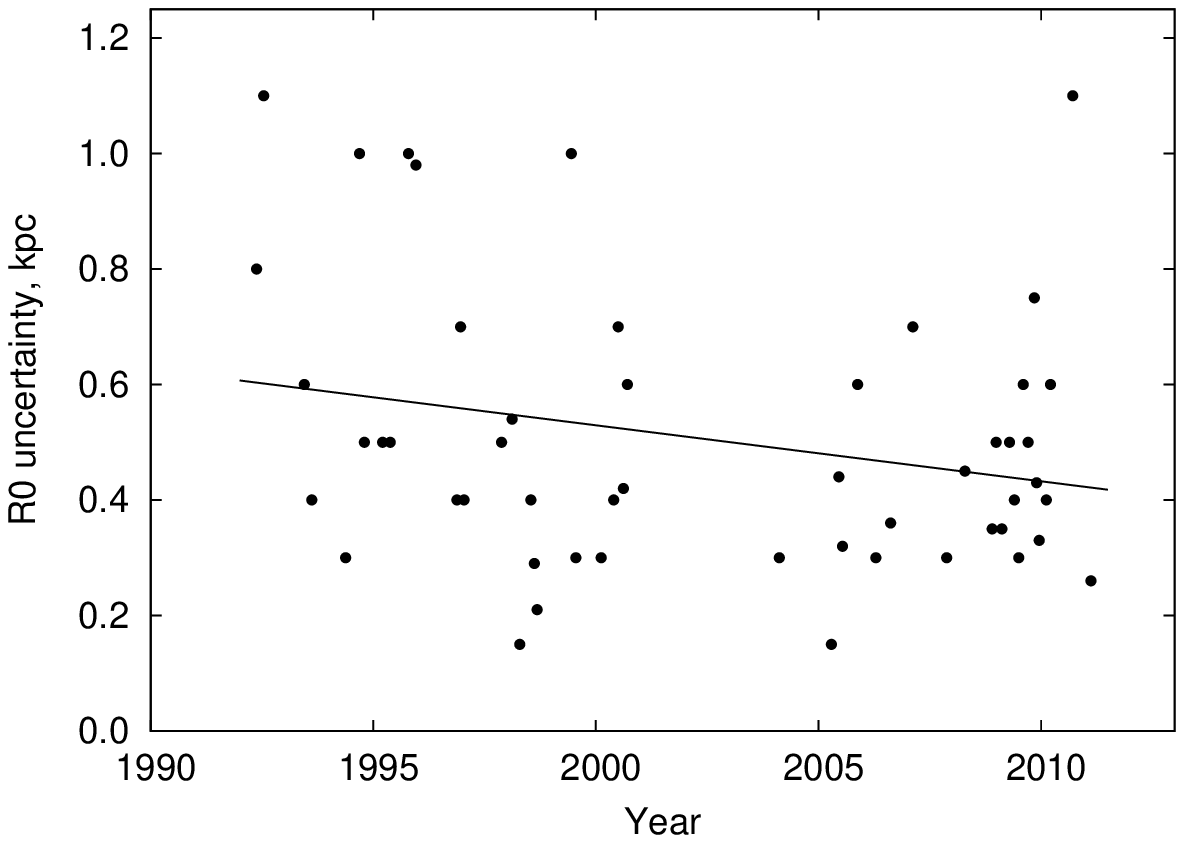}
\caption{Uncertainty in R$_0$ estimates used in this study.}
\label{fig:data_R0_err}
\end{minipage}
\end{figure}

Our analysis used two approaches.  First, we investigated a possible
drift in the R$_0$ values (see Fig.~\ref{fig:data_R0}), which may
indicate the presence of a bandwagon effect.  We obtained a linear
slope in R$_0$ as a function of time of $+0.003 \pm 0.010$ kpc
yr$^{-1}$ for the weighted case and $+0.009 \pm 0.010$ kpc yr$^{-1}$
for the unweighted case. Thus, any trend in R$_0$ estimates published
in the last 20 years is statistically insignificant, which renders a
significant bandwagon effect doubtful. This conclusion is confirmed by
application of the Abbe criterion (Malkin 2013). The latter paper also
contains a detailed discussion on this subject.

On the other hand, the trend in the reported R$_0$ uncertainties (see
Fig.~\ref{fig:data_R0_err}) is statistically significant, i.e.,
$-0.0103 \pm 0.0053$~kpc yr$^{-1}$.  This is particularly interesting
because we expect that there are two tendencies in the changes in the
errors in R$_0$ measurements with time. First, the errors should
decrease with progress in observational and analysis methods.  On the
other hand, one can expect that the published errors should increase
because most authors of recent papers now pay more attention to the
correct computation of the uncertainties in their result. Evidently,
the former tendency prevails.

To assess the internal consistency of the R$_0$ measurements, several
statistical methods were used.  They can be divided into two groups.
The first group consists of computation of the unweighted and
classical weighted means, the author's modification with respect to
computation of the mean uncertainty (Malkin 2001), and three other
weighted-mean modifications, which allow to account for input data
discrepancies, i.e., limitation of relative weights (Nichols 2004),
normalized residuals (James et al. 1992), and the Mandel--Paule method
(Paule \& Mandel 1982).  If all variants of the weighted mean produce
the same mean value and standard error, the input data are consistent.
The second group of methods includes determination of the median with
error estimation (M\"uller 1995) and bootstrap median
methods. Detailed descriptions of these methods are given in Malkin
(2012). The results of the computations are presented in
Table~\ref{tab:all_results}.

\begin{table}[t]
\centering
\caption{Average estimates of R$_0$ obtained with different
  statistical techniques (see text).}
\label{tab:all_results}
\tabcolsep=10pt
\begin{tabular}{lcc}
\hline
\multicolumn{1}{c}{Method} & $R_0$, kpc \\
\hline
Unweighted mean                & $7.979 \pm 0.061$ \\
Classical weighted mean        & $7.967 \pm 0.048$ \\
Modified weighted mean         & $7.967 \pm 0.073$ \\
Limitation of relative weights & $7.967 \pm 0.048$ \\
Normalized residuals           & $7.967 \pm 0.048$ \\
Mandel--Paule method           & $7.958 \pm 0.058$ \\
Median                         & $8.090 \pm 0.062$ \\
Bootstrap median               & $8.060 \pm 0.072$ \\
\hline
\end{tabular}
\end{table}

\section{Concluding remarks}

Although the published R$_0$ estimates were obtained based on
different methods and samples of objects, they are consistent rather
than discrepant.  The results of the computation of a mean R$_0$ value
obtained using different statistical techniques converge at the level
of approximately 0.1~kpc, which is much smaller than the uncertainty
in the average value. Note, however, that this conclusion is not
evident.  Similar analysis of $\Omega_0$ measurements has shown that
the latter are much less consistent.

As significant experience in deriving best estimates of the physical
constants has shown, using various statistical methods to evaluate the
best R$_0$ estimate is very important to assess data consistency and
obtain realistic uncertainties. Therefore, careful astronomical
consideration of the published measurements should be accompanied by a
careful statistical analysis.  It should be recognized that the
computation of the new conventional IAU R$_0$ value is not only an
astronomical, but also a metrological task.

Another result of this study is that any trend in the R$_0$ estimates
obtained during the last 20 years is statistically insignificant,
which makes it unlikely that R$_0$ results are significantly affected
by a bandwagon effect.  On the other hand, the formal errors in the
published R$_0$ estimates improve significantly with time.

Note that the average value R$_0 = 8.0 \pm 0.25$ kpc computed in this
study differs from the results of the latest direct measurements
obtained from stellar orbits about Sgr A*, trigonometric parallaxes to
Sgr B2, and over 60 trigonometric parallaxes of masers, which give
R$_0 = 8.4 \pm 0.2$ kpc (M. J. Reid, priv. comm.), although these
values are still formally consistent.

On the other hand, it seems important to properly combine all results
obtained based on different methods, because this provides better
systematic stability of the average result.  Indeed, the systematic
and random errors of all these results should be assessed and the
corresponding correction should be applied when possible before
averaging (although this is a very difficult task).

\end{document}